\documentclass[a4paper,10pt,pre,twocolumn,showpacs,aps,floats,floatfix,superscriptaddress]{revtex4}

\usepackage{epsfig}
\begin{document}

\newcommand{\avk}{\langle k \rangle}
\newcommand{\fluck}{\langle k^2 \rangle}

\title{Modeling the evolution of weighted  networks}

\author{Alain Barrat} 
\affiliation{Laboratoire de Physique Th\'eorique (UMR du CNRS 8627),
  B\^atiment 210, Universit{\'e} de Paris-Sud 91405 Orsay, France}
\author{Marc Barth\'elemy} \affiliation{CEA-Centre d'Etudes de
  Bruy{\`e}res-le-Ch{\^a}tel, D\'epartement de Physique Th\'eorique et
  Appliqu\'ee BP12, 91680 Bruy\`eres-Le-Ch\^atel, France}
\author{Alessandro Vespignani}
\affiliation{Laboratoire de Physique Th\'eorique (UMR du CNRS 8627),
  B\^atiment 210, Universit{\'e} de Paris-Sud 91405 Orsay, France}


\widetext
\begin{abstract}
  
  We present a general model for the growth of weighted networks in
  which the structural growth is coupled with the edges' weight
  dynamical evolution.  The model is based on a simple weight-driven
  dynamics and a weights' reinforcement mechanism coupled to the local
  network growth. That coupling can be generalized in order to include the
  effect of additional randomness and non-linearities which can be
  present in real-world networks. The model generates weighted graphs
  exhibiting the statistical properties observed in several real-world
  systems. In particular, the model yields a non-trivial time
  evolution of vertices properties and scale-free behavior with
  exponents depending on the microscopic parameters characterizing the
  coupling rules. Very interestingly, the generated graphs
  spontaneously achieve a complex hierarchical architecture
  characterized by clustering and connectivity correlations varying as
  a function of the vertices' degree.
\end{abstract}


\maketitle 

\section{Introduction}

Networked structures appear in a wide array of systems belonging to
domains as diverse as biology, ecology, social sciences, or large
information infrastructures such as the Internet and the World-Wide
Web~\cite{Barabasi:2000,Amaral:2000,mdbook,psvbook}. In recent years,
many empirical findings have uncovered the general occurrence of a
complex topological organization underlying many of these networks,
triggering the interest of the research community. In particular,
small-world properties~\cite{watts98} and large fluctuations in the
connectivity pattern identifying the class of scale-free networks
~\cite{Barabasi:2000} have been repeatedly observed in real world
networks. These findings triggered a wealth of theoretical and
experimental studies devoted to the characterization and modeling of
these features. These studies have pointed out the importance of the
evolution and growth of networks
\cite{Barabasi:1999,mdbook,psvbook} and led to the formulation of a
long list of models aimed at studying the architecture of complex
networks and the dynamical processes which are taking place on their
structure~\cite{havlin00,newman00,barabasi00,pv01a}.

So far the research activity on networks has been mainly focused on
graph in which links are represented as binary states, i.e. either
present or absent. More recently, however, the gathering of more
complete data has allowed to take into account the variation of the
strength of the connections between nodes (i.e. the {\em weights} of
the links), providing a more complete representation of some networked
structures in terms of weighted graphs.  Indeed, this diversity in the
interaction intensity is of crucial interest in real networks. Studies
of congestion phenomena in Internet implies the knowledge and the
characterization of its traffic~\cite{psvbook} and the number of
passengers in the airline networks is obviously a basic information to
assess the importance of an airline
connection~\cite{Amaral:2000,Guimera:2003,Barrat:2004a}. In the case
of ecological networks~\cite{Pimm}, recent studies
(see~\cite{Krause:2003} and references cited) highlighted the
importance of the strength of the predator-prey interaction in
ecosystem stability. Metabolic reactions also carry fluxes that are
essential to the understanding of metabolic networks, as shown
in~\cite{Almaas:2004}.  Finally, sociologists~\cite{Granovetter}
showed already some time ago the importance of weak links in social
networks.  Very interestingly, the analysis of some paradigmatic
weighted networks have revealed that in addition to a complex
topological structure, real networks display a large heterogeneity in
the capacity and intensity of the connections. In particular, broad
distributions and non-trivial correlations between weights and
topology were observed in different
networks~\cite{Guimera:2003,Barrat:2004a,Garla:2003}.

From the previous discussion, it appears clearly that there is a need
for a modeling approach to complex networks that goes beyond the
purely topological point. In this article, we analyze in detail a
general model for the evolution of weighted networks that couples the
topology and weights dynamical evolution.  Vertices entering the
system draw new edges with an attachment dynamics driven by the weight
properties of existing edges and vertices. In addition, in contrast
with previous models~\cite{Yook:2001,zheng03} for which weights are
statically assigned, we allow for the dynamical evolution of weights
during the growth of the system. This dynamics is inspired by the
evolution and reinforcements of interactions in natural and
infrastructure networks.  We provide a detailed analytical and
numerical inspection of the model, considering different specific
mechanisms---homogeneous, heterogeneous, nonlinear---for the 
evolution of weights (A
short report of the simplest linear and homogeneous case appeared in
Ref.~\cite{Barrat:2004b}). The obtained networks display heavy-tailed
distributions of weight, degree and strength. We determine
analytically the exponents of the corresponding power-laws showing
that they depend on the unique parameter defining the model's
dynamics. Interestingly, the model generates graphs that spontaneously
develop a structural organization in which vertices with different
degrees exhibits different level of local clustering and
correlations. These correlations can be shown to emerge as a direct
consequence of the coupling between topology and dynamics. While the
model we introduce here is possibly the simplest one in the class of
weight-driven models, it generates a very rich phenomenology that
captures many of the complex features emerging in the analysis of real
networks. In this perspective it can be considered as a general
starting point for more realistic models aimed at the representation
of specific networks.

The paper is structured as follows. In section~\ref{sec:II} we review
the necessary definitions and tools for the characterization of
complex weighted networks. The general formulation of the model is
reported in section~\ref{sec:III}. Section~\ref{sec:IV} discusses the
homogeneous reinforcement rules and reports the corresponding
analytical and numerical analysis. In Section~\ref{sec:V}, the
dynamics is generalized in order to include the effect of local
randomness. A further generalization to a more complicate non-linear
reinforcement mechanisms for the weights' evolution is discussed and
analyzed in section~\ref{sec:VI}.

\section{Weighted networks}
\label{sec:II}
The topological properties of a graph are fully encoded in its
adjacency matrix $a_{ij}$, whose elements are $1$ if a link connects
node $i$ to node $j$, and $0$ otherwise.  The indices $i,j$ run from
$1$ to $N$ where $N$ is the size of the network and we use the
convention $a_{ii}=0$. Similarly, a weighted network is entirely
described by a matrix $W$ whose entry $w_{ij}$ gives the weight on the
edge connecting the vertices $i$ and $j$ (and $w_{ij}=0$ if the nodes
$i$ and $j$ are not connected). In the following we will consider only
the case of symmetric weights $w_{ij}=w_{ji}$ while the directed case is
considered in~\cite{Digraph}.

Important examples of weighted networks have been recently
characterized. The first example is the world-wide airport network
(WAN)~\cite{Guimera:2003,Barrat:2004a,Li:2003a,Li:2003b} where the weight
$w_{ij}$ is the number of available seats on direct flights
connections between the airports $i$ and $j$. A second important
case-study is the scientific collaboration network
(SCN)~\cite{newmancoll,vicsek} where the nodes are identified with
authors and the weight depends on the number of co-authored
papers~\cite{newmancoll,Barrat:2004a}. These two cases, which are
paradigms of respectively large infrastructure and of social networks,
display complex features characterized by heavy tailed distributions
for topological and weighted quantities. Another very important
example of a weighted network is the biochemical network of metabolic
reactions. For this network, the nodes are biochemical elements
(enzymes, etc) and a link between two nodes denotes the existence of an
individual chemical reaction between them. The weight of a link can be
characterized by the flux of this chemical reaction. A very recent
study~\cite{Almaas:2004} has provided the first analysis of the
weighted graph of a metabolic network, bringing further evidence for
the heterogeneous complex character of weighted networks.

In the following we introduce a set of general quantities whose 
statistical analysis allows the  mathematical characterization 
of the complex and heterogeneous nature of weighted graph.
\subsection{Weights and strength}
The most commonly used topological information about vertices is their
degree and is defined as the number $k_i$ of the neighbors
\begin{equation}
k_i=\sum_{j} a_{ij}\ .
\end{equation}
A natural generalization in the case of weighted networks is the {\em
strength} $s_i$ defined as~\cite{Yook:2001,Barrat:2004a}
\begin{equation}
s_i=\sum_{j}w_{ij} \ .
\end{equation}
Indeed, the strength of a node combines the information about its
connectivity and the intensity of the weights of its links. In the
case of the world-wide airport network, the strength $s_i$ corresponds
to the total traffic going through a vertex $i$ and is therefore an
indication of the importance of the airport $i$. In the case of the
scientific collaboration network, the strength gives the number of
papers authored by a given scientist (excluding single-author
publications, see e.g.~\cite{Barrat:2004a}). 

A natural characterization of the statistical properties of networks
is provided by the probability $P(k)$ that any given vertex has a
degree $k$. Many studies have revealed that networks display a heavy
tailed probability distribution $P(k)$ that in many cases is well
approximated by a power-law behavior $P(k)\sim k^{-\gamma}$ with
$2\leq\gamma\leq 3$.  This has led to the introduction of the class of
scale-free networks~\cite{Barabasi:1999}, as opposed to the regular
graphs with poissonian degree distribution. Similar information on the
statistical properties of weighted networks can be gathered at first
instance by the analysis of the strength and weight distributions
$P(s)$ and $P(w)$ which denote the probability of a vertex to have the
strength $s$ and of a link to have the weight $w$, respectively.  Also
for these distributions, recent measurements on weighted networks have
uncovered the presence of heavy tails and power-law behaviors
~\cite{Guimera:2003,Barrat:2004a,Li:2003a,Li:2003b,Garla:2003,Almaas:2004}.
The heavy-tailed behavior of these distributions is an extremely
relevant characteristic of complex networks indicating the presence of
statistical fluctuations diverging with the graph size. This implies
that the average values $\left\langle k\right\rangle$, $\left\langle
w\right\rangle$, and $\left\langle s\right\rangle$ are not typical in
the network and there is an appreciable probability of finding
vertices with very high degree and strength.  In other words, we are
generally facing networks which are very heterogeneous.  It is worth
stressing that the correlations between the weight and topological
properties are encoded in the statistical relations among these
quantities. Indeed, $s_i$, which is a sum over all neighbors of $i$,
is correlated with its degree $k_i$. In the simplest case of random,
uncorrelated weights $w_{ij}$ with average $<w>$, the strength is
$s\sim <w>k$. In the presence of correlations between weights and
topology, we may observe a more complicated behavior with $s\sim A
k^{\beta}$ with $\beta \ne 1$ or with $\beta=1$ and $A \ne <w>$.

\subsection{ Clustering and correlation}
Complex networks display an architecture imposed by the structural and
administrative organization of these systems that is not fully
characterized by the distributions $P(k)$ and $P(s)$. Indeed, the
structural organization of complex networks is mathematically encoded
in the various correlations existing among the properties of different
vertices.  For this reason, a set of topological and weighted
quantities are customarily studied in order to uncover the network
architecture.  A first and widely used quantity is given by the {\em
clustering} of vertices.  The clustering of a vertex $i$ is defined as
\begin{equation}
c_i = \frac{1}{k_i (k_i -1)} \sum_{j,h} a_{ij}a_{ih}a_{jh} \ ,
\end{equation}
and measures the local cohesiveness of the network in the neighborhood of
the  vertex. Indeed, it yields  the fraction of inter-connected 
neighbors of a given vertex. The average over all vertices gives the 
network {\em clustering coefficient} which describes the statistics 
of the density of connected triples. Further information can be gathered by
inspecting the average clustering coefficient $C(k)$ restricted to
classes of vertices with degree $k$:
\begin{equation}
C(k) = \frac{1}{NP(k)} \sum_{i/k_i=k} c_i \ .
\end{equation}
In many networks, the degree-dependent clustering coefficient $C(k)$
is a decreasing function of $k$ which shows that low-degree nodes
generically belong to well interconnected communities while
high-degree sites are linked to many nodes that may belong to
different groups which are not directly
connected~\cite{Vazquez02,Ravasz02}. This is generally the signature
of a non trivial architecture in which hubs, high degree vertices,
play a distinct role in the network. 

Another important source of information lies in the correlations of the
degree of neighboring vertices~\cite{Pastor-Satorras:2001,Maslov:2001}. 
Since the whole conditional distribution $P(k'|k)$ that a 
given site with degree $k$ is connected to another site of 
degree $k'$ is often difficult to interpret, the {\em
average nearest neighbor degree} has been proposed to measure these
correlations~\cite{Pastor-Satorras:2001}
\begin{equation}
k_{nn,i}=\frac{1}{k_i}\sum_{j=1}^N a_{ij} k_j \ .
\end{equation}
Once averaged over classes of vertices with connectivity $k$, the 
average nearest neighbor degree can be expressed as 
\begin{equation}
k_{nn}(k) = \sum_{k'} k' P(k'|k) \ ,
\end{equation}
providing a probe on the degree correlation function.
If degrees of neighboring vertices are uncorrelated, $P(k'|k)$ is only
a function of $k'$ and thus $k_{nn}(k)$ is a constant. When
correlations are present, two main classes of possible correlations
have been identified: {\em assortative} behavior if $k_{nn}(k)$
increases with $k$, which indicates that large degree vertices are
preferentially connected with other large degree vertices, and {\em
disassortative} if $k_{nn}(k)$ decreases with $k$~\cite{Newman:2002}.

\begin{figure}[t]
\vskip .5cm
\begin{center}
\epsfig{file=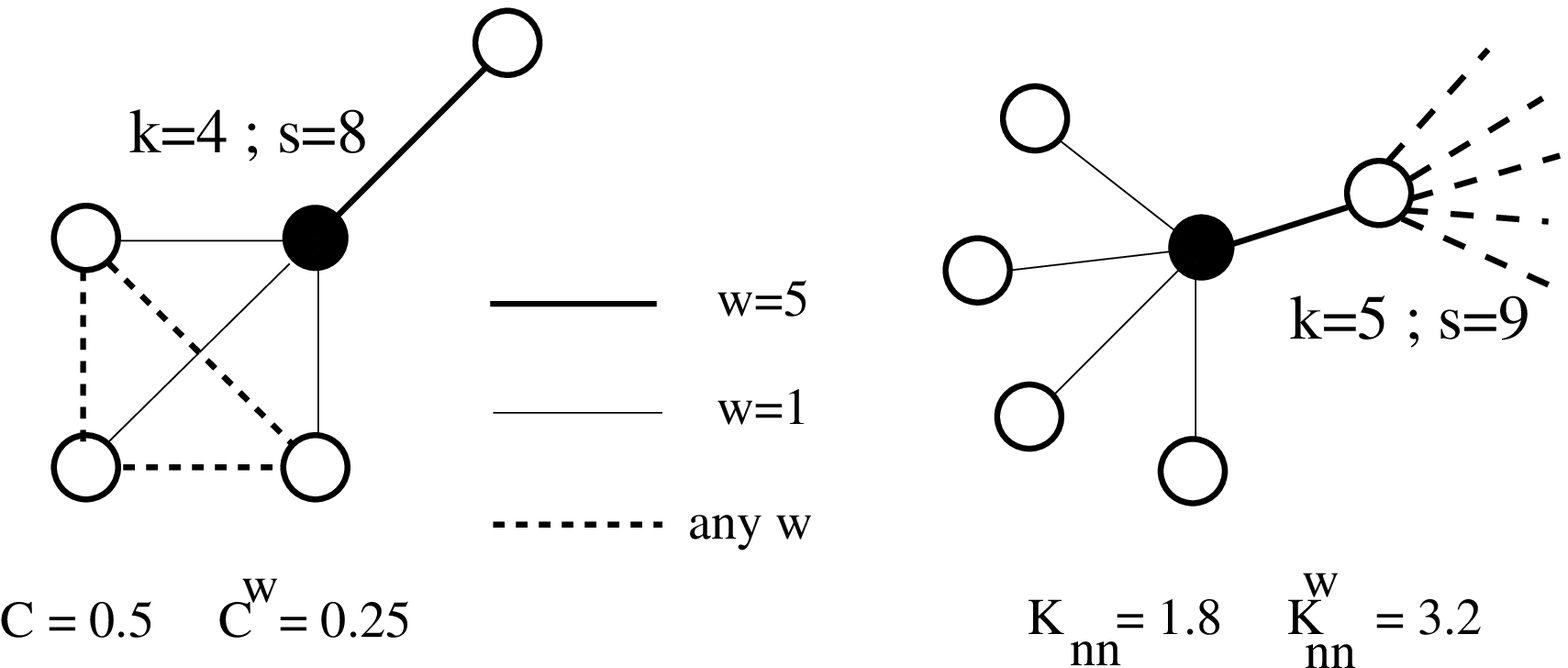,width=6cm}
\end{center}
\begin{center}
\epsfig{file=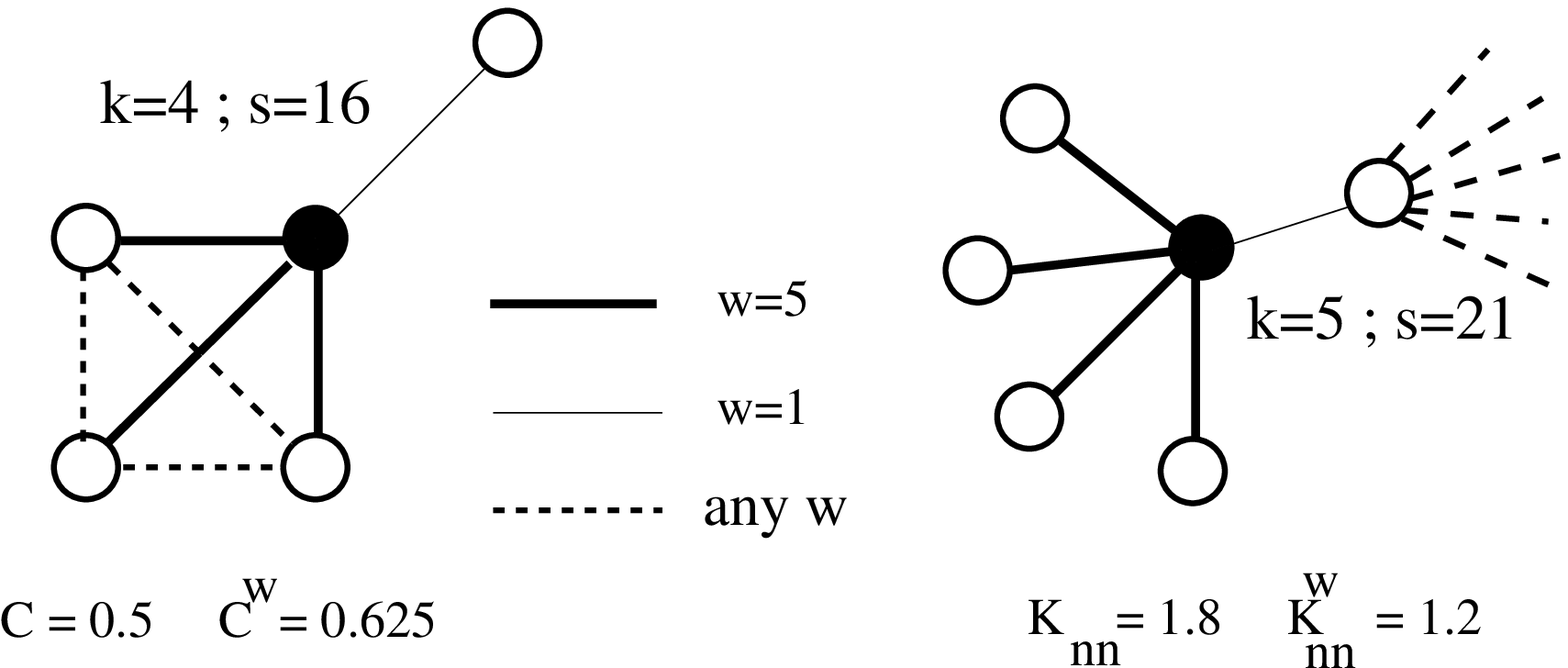,width=6cm}
\end{center}
\caption{Examples of local configurations whose topological and
weighted quantities are different. Top: In both cases the central
vertex (filled) has a very strong link with only one of its
neighbors. Bottom: opposite situation.  The weighted clustering and
the weighted average nearest neighbors degree capture more precisely
than their topological counterparts the effective level of
cohesiveness and affinity due to the actual interaction intensity as
measured by the weights.}
\label{figcwknnw}
\end{figure}

While the above quantities provide clear signatures of the structural
organization, they are defined solely on topological grounds and the
inclusion of weights and their correlations might be extremely
important for a full understanding of the networks' architecture.  For
instance, Fig.~\ref{figcwknnw} clearly shows that very different
situations in terms of weights can have the same topological
clustering: if the existing triples are formed by links with small
weights, the (geometrical) clustering coefficient will overestimate
their relevance in the network's organization.  For this reason,
generalizations of clustering and correlations measurements to
weighted networks have been put forward in~\cite{Barrat:2004a}.

The {\em weighted clustering
coefficient} of a vertex is defined as~\cite{Barrat:2004a}
\begin{equation}
  c_{i}^w=\frac{1}{s_i(k_i-1)} \sum_{j, h}
  \frac{(w_{ij}+w_{ih})}{2} a_{ij}a_{ih}a_{jh}.
\end{equation}
This quantity combines the measure of the existence of triples around
vertex $i$ with the intensity of the links emanating from $i$ and
participating to these triples. As we show in Fig.~\ref{figcwknnw},
$c_{i}^w$ describes more accurately than $c_i$ the relevance of these
triples. The normalization $s_i(k_i-1)$ corresponds to the maximum
possible value of the numerator and thus ensures that $c_i^w \in
[0,1]$. The average over all sites, or over sites of a given degree
$k$, define respectively the global weighted clustering coefficient
$C^w$ and $C^w(k)$.  For random or uniform weights, these averages
coincide with their geometrical counterparts.  On the other hand, the
comparison between $C$ and $C^w$ (and also between $C(k)$ and
$C^w(k)$) conveys informations on the repartition of weights. A larger
weighted clustering $C^w > C$ signals that links with large weights
have a tendency to form triples while the opposite case $C^w < C$
signals a lower relevance of the triangles.

Analogously, high degree vertices could be connected mainly to
small degree vertices with a small intensity of connections and to few
large degree vertices with large weights: a topological
disassortative character is therefore emerging while in term of 
interactions one would conclude to an assortative behavior. In the same
spirit as for the clustering, one can therefore define 
the {\em weighted average nearest neighbor degree} as~\cite{Barrat:2004a}
\begin{equation}
  k^w_{nn,i}=\frac{1}{s_i}\sum_{j=1}^N w_{ij} k_j \ .
\end{equation}
This quantity is the natural generalization of the usual assortativity
$k_{nn,i}$ and balances the nearest neighbor degree with the
normalized weight of the connecting edge $w_{ij}/s_i$. It reduces to
$k_{nn,i}$ for uniform or random weights.  Comparing $k^w_{nn,i}$ with
$k_{nn,i}$ informs us if the larger weights point to the neighbors
with larger degree (if $k^w_{nn,i} > k_{nn,i}$) or on the contrary to
the ones with smaller degree (see Fig.~\ref{figcwknnw}). The
behavior of $k_{nn}^w(k)$, thus measures the effective {\em affinity}
to connect with high or low degree neighbors according to the
magnitude of the actual interactions.

In the following we will make use of all these quantities in order to
provide a throughout characterization of the weighted graphs generated
by our model and to assess the relevance of the weights in
their structural organization.  

\section{The model}
\label{sec:III}

Previous models  of weighted growing
networks~\cite{Yook:2001,zheng03} were considering  
the growth as driven by the topological features only,
with weights statically assigned to the links; i.e. 
$w_{ij}$ is chosen at the creation of the link $i-j$ and does not evolve
afterwards. This mechanism leads to topological heterogeneities,
however it lacks any dynamical feature of the network's weights.  
Indeed, it is rather intuitive to consider that the addition of new
vertices and  links will perturb, at least locally, the existing
weights. This phenomenon can be easily understood by considering  
the example of the airline network: a new airline connection arriving 
at airport A will generally modify (increase) the traffic activity 
between airport A and its neighboring airports. Passengers brought by
the new connection will eventually get on connection flights,
increasing the passenger flow on the other routes. In the Internet as
well, it is easy to realize that the introduction of a new connection
to a router corresponds to an increase in the traffic handled on the
other router's links. Indeed in many technological, large infrastructure
and social networks we are generally led to think about a
reinforcement of the weights due to the network's growth.
In this spirit we consider here a model for growing weighted network 
that takes into account the coupled evolution in time of topology and 
weights~\cite{Barrat:2004b} and leaves room for accommodating different
mechanisms for the reinforcement of interactions. 

The definition of the model is based on two coupled mechanisms: the 
topological growth and the weights' dynamics.

(i){ \it Growth}. Starting from  an initial seed of $N_0$ vertices
connected by links with assigned weight $w_0$, a
new vertex $n$ is added at each time step. This new site is connected
to $m$ previously existing vertices, choosing preferentially sites with
large strength; i.e. a node $i$ is chosen according to the probability
\begin{equation}
\Pi_{n\to i}=\frac{s_i}{\sum_j s_j}.
\label{sdrive}
\end{equation}
This rule, of {\em strength driven attachment}, generalizes the usual
preferential attachment mechanism driven by the topology, to weighted
networks. Here, new vertices connect more likely to vertices which are
more central in terms of the strength of interactions.

\begin{figure}[t]
\vskip .5cm
\begin{center}
\epsfig{file=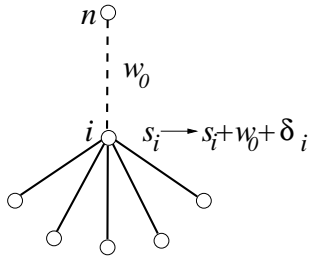,width=4cm}
\end{center}
\caption{ Illustration of the construction rule. A new node $n$
connects to a node $i$ with probability proportional to 
$ s_i/\sum_j s_j$. The weight of the new edge is $w_0$ and the 
total weight on the existing edges connected to $i$ is modified 
by an amount equal to $\delta_i$. }
\label{fig:rule}
\end{figure}

(ii){\it Weights' dynamics}. The weight of each new edge $(n,i)$ is
initially set to a given value $w_0$.  
The creation of this edge will introduce variations of the
traffic across the network. For the sake of simplicity we limit
ourselves to the case where the introduction of a new edge on
node $i$ will trigger only local rearrangements of weights on the
existing neighbors $j\in{\cal V}(i)$, according to the rule 
\begin{equation}
w_{ij}\to w_{ij}+\Delta w_{ij} \ ,
\label{rule}
\end{equation}
where in general $\Delta w_{ij}$ depends on the local dynamics and can
be a function of different parameters such as the weight $w_{ij}$, the
connectivity or the strength of $i$, etc. In the following we focus
on the case where the addition of a new edge with weight $w_0$ induces
a total increase $\delta_i$ of the total outgoing traffic and where
this perturbation is proportionally distributed among the edges
according to their weights [see Fig.~(\ref{fig:rule})]
\begin{equation}
\Delta w_{ij}=\delta_i \frac{w_{ij}}{s_{i}} \ .
\label{rulep}
\end{equation}
This rule yields a total strength
increase for node $i$ of $\delta_i+w_0$, implying that 
$s_i\to s_i+\delta_i+w_0$. 
After the
weights have been updated, the growth process is iterated by
introducing a new vertex, i.e. going back to step (i) until the
desired size of the network is reached.

The mechanisms (i) and (ii) have  simple physical and realistic
interpretations. Equation~(\ref{sdrive}) corresponds to the fact that
new sites try to connect to existing vertices with the largest
strength. This is a plausible mechanism in many real world networks.
For instance, in the Internet new routers connect to
routers that have larger bandwidth and traffic handling
capabilities. In the case of the airport's networks, new connections
are generally established to airports with a large passenger
traffic. In contrast to the connectivity preferential attachment of
the ``rich get richer'' type, the mechanism here relies on the
importance of the traffic and could be more adequately described as
``busy get busier''.
At the same time, the weights' dynamics
Eqs.~(\ref{rule},\ref{rulep}) 
couples the  addition of new edges and vertices with the 
evolution of weight and strength and correspond
to different scenarios according to the value of $\delta_i$:
\begin{itemize}
\item for $\delta_i <w_0$, the new link has not a large
influence. This may be the case for scientific collaborations where
the birth of a new collaboration (co-authorship) is very likely not
going to strengthen the activity on previous collaborations.
\item $\delta_i \approx w_0$ corresponds to situations for which the
new created traffic (on the new link $n-i$) is transferred onto the
already existing connections in a ``conservative'' way.
\item $\delta_i >w_0$ is an extreme case in which a new edge generates
a sort of multiplicative effect that is bursting the weight or traffic
on neighbors.
\end{itemize}
The quantity $w_0$ sets the scale of the weights and we can therefore
use the rescaled quantities $w_{ij}/w_0$, $s_i/w_0$ and
$\delta_i/w_0$, or equivalently set $w_0=1$. The model then depends
only on the dimensionless parameter $\delta_i$. The generalization to
arbitrary $w_0$ is simply obtained by replacing $\delta_i$, $w_{ij}$
and $s_i$ respectively by $\delta_i/w_0$, $w_{ij}/w_0$ and $s_i/w_0$
in all results.

The model is very general and the properties obtained for the
generated networks will strongly depend on the kind of coupling
between topology and weights as specified by the parameter $\delta_i$
and its variations depending on the vertex' properties. In the
following we will provide analytical and numerical inspections of
three prototypical situations that can be used as starting points for
further generalizations.

\section{Homogeneous coupling}
\label{sec:IV}
In this section, we will focus on the simplest form of coupling with
$\delta_i=\delta=const$. This case amounts to a very homogeneous
system in which all the vertices have an identical coupling between
the addition of new edges and the corresponding weights'
increase. Such a growing model can be analytically studied through the
time evolution of the {\em average} value of $s_i(t)$ and $k_i(t)$ of
the $i$-th vertex at time $t$, neglecting fluctuations and thus
working at a ``mean-field'' level.  In addition, we use numerical
simulations in order to provide a direct statistical analysis
of the generated graph and substantiate the analytical findings.

\subsection{Evolution of strength and weights}

The network growth starts from an initial seed of $N_0$ nodes, and
continues with the addition of one node per unit time, until a size
$N$ is reached. When a new edge $n$ is added to the network, an
already present vertex $i$ can be affected in two ways: i) It is
chosen with probability (\ref{sdrive}) to be connected to $n$; then
its connectivity increases by $1$, and its strength by $1+\delta$.
ii) One of its neighbors $j \in V(i)$ is chosen to be connected to
$n$. Then the connectivity of $i$ is not modified but $w_{ij}$ is
increased according to the rule Eq.~(\ref{rule}), and thus $s_i$ is
increased by $\delta w_{ij}/s_j$. This dynamical process modulated by
the respective occurrence probabilities $s_i(t)/\sum_l s_l(t)$ and
$s_j(t)/\sum_l s_l(t)$ is thus described by the following evolution
equations for $s_i$ and $k_i$
\begin{eqnarray}
\nonumber
\frac{ds_i}{dt}&=&m\frac{s_i(t)}{\sum_l s_l(t)}(1+\delta)+
\sum_{j\in{\cal V}(i)}m\frac{s_j(t)}{\sum_l s_l(t)}
\delta\frac{w_{ij}(t)}{s_j(t)}\\
\frac{dk_i}{dt}
&=&m\frac{s_i(t)}{\sum_l s_l(t)} \ ,
\label{eq_evol_sk}
\end{eqnarray}
where we have considered the continuous approximation that treats $k$,
$s$ and the time $t$ as continuous
variables~\cite{Barabasi:2000,mdbook}.  These equations may be written
in a more compact form by noticing that the addition of a node results
in the addition of $m$ links obtaining that the total degree at time
$t$ is given by $\sum_{i=1}^t k_i(t) \approx 2 m t $.  Similarly, each
added link increases the total strength by an amount equal to
$2+2\delta$, obtaining
\begin{equation}
\sum_{i=1}^t  s_i(t)\approx 2m(1+\delta)t \ .
\end{equation}
By plugging this result into the equations (\ref{eq_evol_sk}), we
obtain the following dynamical equations
\begin{eqnarray}
\nonumber
\frac{ds_i}{dt}&=&
\frac{2\delta +1}{2\delta +2} \frac{s_i(t)}{t}\\
\frac{dk_i}{dt}
&=&\frac{s_i(t)}{2 (1+\delta) t} \ .
\label{eq_evol2}
\end{eqnarray}
These equations can be readily integrated with  
initial conditions $k_i(t=i)=s_i(t=i)=m$, yielding
\begin{eqnarray}
\label{eq_ss.vs.t}
s_i(t)=m ~\left(\frac{t}{i}\right)^{\frac{2\delta+1}{2\delta+2}} \\
k_i(t)=\frac{s_i(t) + 2m\delta}{2\delta+1}.
\label{eq_sk.vs.t}
\end{eqnarray}
\begin{figure}
\vskip .5cm
\begin{center}
\epsfig{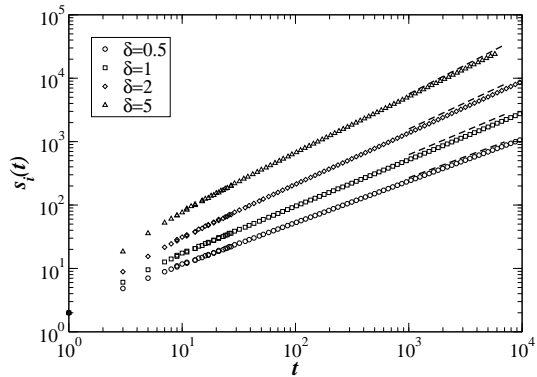}
\end{center}
\caption{ Evolution of the strength of vertices during the growth
of the network, for various values of $\delta$; The thick lines
are the predicted power laws $t^a$, $a=(1+2\delta)/(2+2\delta)$ 
($m=2$, $N=10^4$). 
}
\label{fig:st}
\end{figure}
The strength and degree of vertices are thus related by the following 
expression
\begin{equation}
s_i = (2\delta +1)k_i - 2m\delta
\label{eq:s_vs_k}
\end{equation}
that implies a proportionality between strength and degree.
It is worth noticing, however, that this relation indicates  
the existence of correlations that are not present in the case of randomly
assigned weights. Indeed, at each new link created the sum of weights
is incremented by $1+\delta$ and therefore $<w> = 1+\delta$. As
previously mentioned, a random assignment of weights would then lead
to $s_i = k_i (1+\delta)$. The equation (\ref{eq:s_vs_k}) instead reveals
a different proportionality constant signaling correlations
between the two quantities. 
The proportionality relation $s\sim k$ also
indicates that the weight-driven dynamics generates in
Eq.~(\ref{sdrive}) an effective degree preferential attachment. This
model thus displays a microscopic mechanism accounting for the
presence of the preferential attachment dynamics in growing networks.

\begin{figure}[htb]
\vskip .5cm
\begin{center}
\epsfig{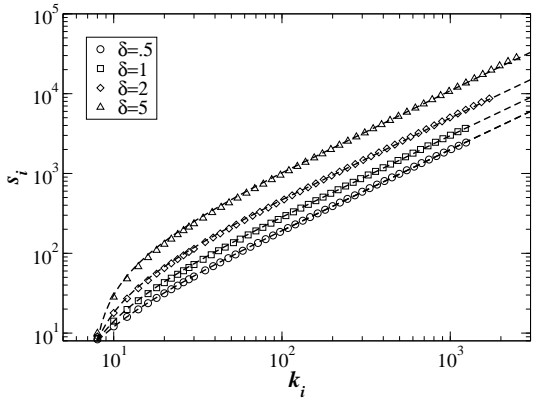}
\end{center}
\caption{Strength $s_i$ versus $k_i$ for $m=8$ and for various values
of $\delta$ ($N=10^4$, data averaged over $1000$ samples). The dashed
lines are the predictions of Eq.~(\ref{eq:s_vs_k}):
$s_i=(1+2\delta)k_i - 2 m \delta$.  }
\label{fig:s_vs_k}
\end{figure}

In order to check the analytical predictions we performed numerical
simulations of networks generated by using the present model with 
different values of $\delta$, minimum degree $m$ and varying network 
size $N$. In Fig.~\ref{fig:st} we show the behavior of the vertices' 
strength versus time for different values of $\delta$, recovering the 
behavior predicted analytically. 
We also report the average strength $s_i$ of vertices 
with degree $k_i$  and confirm that $\beta=1$ as well as the
validity of Eq. (\ref{eq:s_vs_k}) as shown in Fig.~\ref{fig:s_vs_k}.

The time evolution of the weights $w_{ij}$ can also be computed
analytically along the lines used for the study of $s_i(t)$ and
$k_i(t)$. 
Indeed, $w_{ij}$ evolves each time a new node connects to either $i$
or $j$ and the corresponding
evolution  equation can  be written as 
\begin{eqnarray}\nonumber
\frac{dw_{ij}}{dt}&=&m\frac{s_i}{\sum_ls_l}\delta\frac{w_{ij}}{s_i}
+m\frac{s_j}{\sum_ls_l}\delta\frac{w_{ij}}{s_j}\\
&=&\frac{\delta}{2(1+\delta)}
\frac{w_{ij}}{t} \ .
\end{eqnarray}
The link $(i,j)$ is created at $t_{ij}=\max(i,j)$ with initial condition
$w_{ij}(t_{ij})=1$, so that
\begin{equation}
w_{ij}(t)=\left(\frac{t}{t_{ij}}\right)^{\frac{\delta}{\delta+1}}
\label{w_t}
\end{equation}
At fixed time $t$, this result implies that
\begin{equation}
w_{ij}(t)\sim\min(i,j)^{\frac{\delta}{\delta+1}}\ .
\end{equation}
and since $k_i(t) \sim {i}^{-\frac{2\delta+1}{2\delta+2}}$, we obtain
\begin{equation}
w_{ij} \sim \min(k_i,k_j)^{\frac{2\delta}{2\delta+1}}\ .
\label{w_k}
\end{equation}

\begin{figure}[htb]
\vskip .5cm
\begin{center}
\epsfig{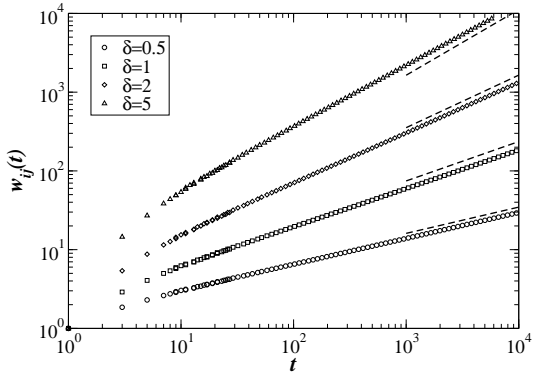}
\end{center}
\caption{Time evolution of $w_{ij}$ during the
growth of the network, for different values of $\delta$.
The functional behavior is consistent with the predicted power law $t^b$, 
$b=\delta/(1+\delta)$, shown as dashed lines. 
Data are averaged over $200$ networks with $m=2$ and  
$N=10^4$.
}
\label{fig:wt}
\end{figure}

\begin{figure}[htb]
\vskip .5cm
\begin{center}
\epsfig{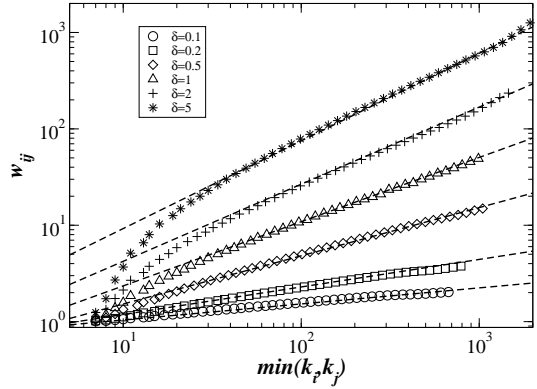}
\end{center}
\caption{$w_{ij}$ vs. $\min(k_i,k_j)$; the dashed lines are the 
theoretical predictions ($m=8$, $N=10^4$, data averaged over $1000$
samples).  }
\label{fig:wvsmin_m2}
\end{figure}

In this case also, the numerical simulations of the model reproduce
the behaviors predicted by the analytical calculations. The time
evolution of some randomly chosen weights is displayed in
Fig.~\ref{fig:wt} and compared with the prediction of
Eq.~(\ref{w_t}). For the sake of completeness we show in
Fig.~\ref{fig:wvsmin_m2} the validity of Eq.~(\ref{w_k}) for the
correlation between $w_{ij}$ and the connectivities of $i$ and $j$.

\subsection{Probability distributions}

\begin{figure}[t]
\vskip .5cm
\begin{center}
\epsfig{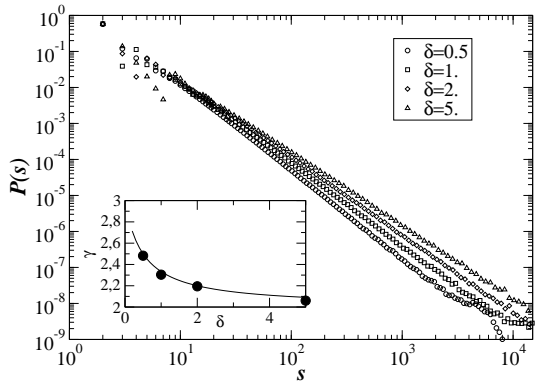}
\end{center}
\caption{Probability distribution $P(s)$.  Data are consistent with a
power-law behavior $s^{-\gamma}$.  In the inset we report the value of
$\gamma$ obtained by data fitting (filled circles), together with the
analytic expression $\gamma=(3+4\delta)/(1+2\delta)$ (line).  The data
are averaged over $200$ networks of size $N=10^5$.  }
\label{fig:P_s}
\end{figure}

The knowledge of the time-evolution of the various quantities
allows us to compute their statistical properties. 
Indeed, the time $t_i=i$ at
which the node $i$ enters the network is uniformly distributed in
[0,t] and the degree probability distribution can be written as
\begin{equation}
P(k, t) = \frac{1}{t+N_0} \int_0^{t} \delta(k - k_i(t)) dt_i, 
\label{eqdistr}
\end{equation} 
where $\delta(x)$ is the Dirac delta function. Using equation $k_i(t)
\sim (t/i)^a$ obtained from Eqs.~(\ref{eq_ss.vs.t},\ref{eq_sk.vs.t})
one obtains in the infinite size limit $t\to\infty$
the distribution $P(k)\sim k^{-\gamma}$ with $\gamma=1+1/a$:
\begin{equation}
\gamma=\frac{4\delta+3}{2\delta+1}.
\end{equation}
Since $s$ and $k$ are proportional, the same behavior $P(s)\sim
s^{-\gamma}$ is obtained for the strength distribution.  $P(s)$ is
displayed in Fig.~\ref{fig:P_s}, showing that the obtained graph is a
scale-free network both for topology and strength and described by an
exponent $\gamma\in[2,3]$ that depends on the value of the parameter
$\delta$. As expected, if the addition of a new edge does not affect
the existing weights ($\delta=0$), we recover the Barab\'asi-Albert
model~\cite{Barabasi:1999} with the value $\gamma=3$. As $\delta$
increases, the distributions get broader with $\gamma\to 2$ when
$\delta\to \infty$, i.e. in a range of values usually observed in the
empirical analysis of networked
structures~\cite{Barabasi:2000,mdbook,psvbook}. This result could be
an explanation of the lack in real-world networks of any universality
of the degree distribution exponent. Our model indeed predicts that
all the exponents will be non-universal and depend on the local
processes which take place at nodes receiving new links.

\begin{figure}[t]
\vskip .5cm
\begin{center}
\epsfig{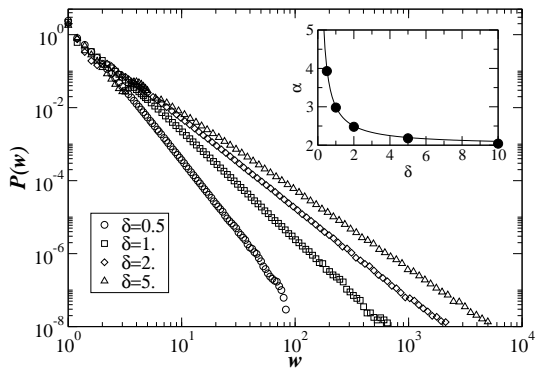}
\end{center}
\caption{ Probability distribution of the weights $P(w)\sim
w^{-\alpha}$.  In the inset we report the value of $\alpha$ obtained
by data fitting (filled circles) and the analytic expression
$\alpha=2+1/\delta$ (solid line).  The data are averaged over $200$
networks of size $N=10^5$.}
\label{fig:P_w}
\end{figure}

The weight distribution $P(w)$ can be analogously calculated 
yielding the power-law behavior
\begin{equation}
P(w)\sim w^{-\alpha},\ \ \ \ \alpha=2+\frac{1}{\delta} \ .
\end{equation}
The distribution $P(w)$ therefore is even more sensitive to the
parameter $\delta$ and evolves from a delta function for
$\delta=0$ (no evolution of the weights) to a very broad power-law as
$\delta\to\infty$. The value $\delta=1$ corresponds to $\alpha=3$,
i.e. to the boundary between finite and unbounded fluctuations of the
weights. In Fig.~\ref{fig:P_w}, 
the weight distributions obtained from numerical simulations at
different values of the parameter $\delta$ are reported along with the
comparison between the values of the measured exponent $\alpha$ and the 
analytical prediction. The precise microscopic dynamics ruling the 
network's growth and the rearrangement of weights is therefore very 
relevant to the final distribution of weights, even if it affects 
only in a much milder way degree and strength. 

\subsection{Correlations and clustering}

The previous analytical study of the model does not provide
information on the correlations generated by the growing process. In
order to have a direct inspection of these properties we therefore
consider the graphs generated by the model for different values of
$\delta$, $m$ and $N$ and measure the quantities defined in
section~\ref{sec:II}, that characterize the clustering and correlation
properties.

\begin{figure}[htb]
\vskip .5cm
\begin{center}
\epsfig{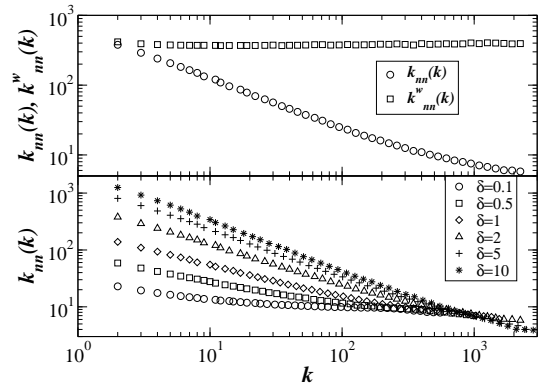}
\end{center}
\caption{$m=2$; Top: $k_{nn}(k)$ and $k_{nn}^w(k)$ for $\delta=2$; bottom:
$m=2$; evolution of $k_{nn}(k)$ for increasing $\delta$.
}
\label{fig:knn_knnw_m2}
\end{figure}

The model exhibits also in this case properties which are depending on
the basic parameter $\delta$. More precisely, for small $\delta$, the
average nearest neighbor degree $k_{nn}(k)$ is quite flat as in the BA
model. The disassortative character emerges as $\delta$ increases and
gives rise to a power law behavior of $k_{nn}(k) \sim k^{-a}$ as shown
in Fig.~\ref{fig:knn_knnw_m2}~\cite{Noteknn}. Remarkably, the weighted
average nearest neighbor degree displays for any $\delta$ a flat
behavior. Moreover, as in recently studied real networks, $k_{nn}^w(k)
> k_{nn}(k)$ which indicates that the larger weights contribute to the
links towards vertices with larger connectivities. This behavior, also
obtained in real weighted networks, is shown in
Fig.~\ref{fig:knn_knnw_m2}.
\vskip .5cm

\begin{figure}[htb]
\vskip .5cm
\begin{center}
\epsfig{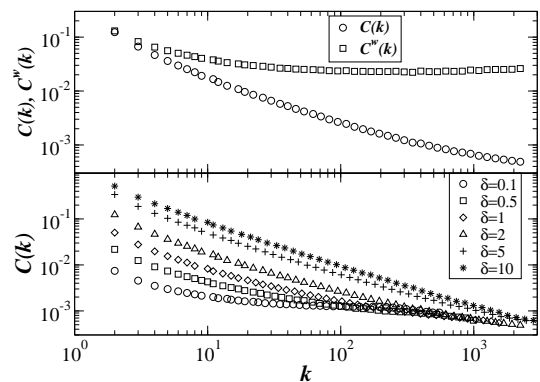}
\end{center}
\caption{$m=2$; Top: $C(k)$, $C^w(k)$ for $\delta=2$;
bottom: $C(k)$ for various $\delta$.
}
\label{fig:ck_ckw_m2}
\end{figure}

Analogous properties are obtained for the clustering spectrum.
At small $\delta$, the clustering coefficient of the network is small
and $C(k)$ is flat. As $\delta$ increases
however, the global clustering coefficient increases and $C(k)$
becomes a decreasing power-law similar to real networks data~\cite{Noteck}.
Fig.~(\ref{fig:ck_ckw_m2}) clearly shows that the increase in
clustering is determined by small k vertices.
The weighted clustering $C^w$ also increases and is larger than the
topological $C$, with an essentially flat $C^w(k)$. Especially at
large $k$, it is clear that the usual clustering coefficient
underestimates the importance of triples in the network since, for the
hubs, the edges with the highest weights belong in great part to the
interconnected triples. In other words, interconnected vertices are
joined by edges that have weights larger than the average value found
in the network.

Interestingly, correlations and clustering spectrum can be qualitatively
understood by considering the dynamical process leading to the 
formation of the network. Indeed, vertices with 
large connectivities and strengths are the ones that entered the system
at the early times as shown by equations
(\ref{eq_ss.vs.t},\ref{eq_sk.vs.t}). Newly arriving vertices attach to 
pre-existing vertices with large strength which on their turn are 
reinforced by the rearrangement of the weights. 
This process naturally builds up a hierarchy
among the nodes: ``old'' vertices have neighbors that are more likely to 
be ``young'' i.e. with small connectivity. On the contrary, newly 
arriving vertices on the contrary have neighbors with high degree and 
strength, generally leading to a disassortative behavior. 
This effect gets stronger as $\delta$ increases and $P(k)$ broadens 
leading to disassortative properties. 
As well, edges among ``old'' vertices are the ones that gets more 
reinforced by the weights dynamics indicating that 
the edges between older nodes, with large
connectivities will be typically stronger than the average. 
This means that the weighted
assortativity will be larger than the topological assortativity,
leading to $k_{nn}^w(k) > k_{nn}(k)$, especially for large $k$.

Similarly, the increase of $C$ with $\delta$ is also directly related
to the mechanism which rearranges the weights after the addition of a
new edge. Since vertices with large strength and degree are generally
connected among them, a new vertex has more probability to attach to
the extremities of a given edge. Triangles will typically be made of
two ``old'' nodes and a ``young'' one. therefore $C(k)$ increases
faster for ``younger'' (low degree) nodes when $\delta$ increases
generating the observed spectrum. On the other hand, the edges between
``old'' and ``young'' vertices are the most recent ones and do not
have large weights. This feature implies that for low degree vertices
$c_i$ and $c_i^w$ are rather close. In contrast, high degree vertices
are connected to each other by edges with large weights, leading to a
weighted clustering coefficient larger than the topological one.

These qualitative arguments confirm the importance of considering weighted
correlations since topological correlations do not fully reveal the
intrinsic coupling between topology and weights, that may lead to very
 different behavior of the correlation and clustering spectrum. 

\section{heterogeneous coupling}
\label{sec:V}

In the model described in the previous section as well as in most
models of growing networks, connectivities and strengths of different
vertices grow with the same exponent (see
Eq.~(\ref{eq_sk.vs.t})). Therefore, vertices entering the system at
the early times have always the largest connectivities and
strengths. One can however imagine that a newly arriving vertex has
intrinsic properties which make it more attractive than older ones so
that its connectivity and strength could grow faster than its
predecessors. This feature is certainly very important in many real
systems where individuals are not identical and has been put forward
in the so-called fitness model \cite{Bianconi:2001} [see also
\cite{Caldarelli:2002} for a static definition of a fitness model].
In a very similar spirit, we introduce here a node-dependent
$\delta_i$ implying that the perturbation of weights created by any
new edge attached to the node $i$ will depend on the very local
properties of $i$. This amounts to introduce a general heterogeneity
in the dynamical properties of the elements of the system.

In this case, each vertex entering the system is tagged with its own 
$\delta_i$ that we will assume are independent
random variables taken from a given distribution $\rho(\delta)$ 
characterizing the system's heterogeneity.
The preferential attachment (Eq.~\ref{sdrive}) 
is not modified and the redistribution of weights now reads 
\begin{equation}
\Delta w_{ij} = \delta_i \frac{w_{ij}}{s_i} \ .
\end{equation}
A large value of $\delta_i$ does not favor immediately the
attractiveness of the vertex but the addition of a new link to $i$
modifies its total strength by a large amount
\begin{equation}
s_i \rightarrow s_i + w_0 + \delta_i.
\end{equation}
On the long run, larger $\delta_i$ yield therefore larger 
increases $\Delta s_i$ when $i$ is chosen for the
addition of a new edge. Since the model's dynamics is driven by 
a strength driven attachment the vertices with larger $\delta_i$ 
will be progressively  favored in the establishment of new
connections, therefore achieving a faster degree and strength growth
as time goes by. 

Similarly to the homogeneous model of the previous section, the 
evolution equations of $s_i$ and $k_i$ can be  written as
\begin{eqnarray}\nonumber
\label{eq_evol_ss_rand}
\frac{ds_i}{dt}&=&m\frac{s_i(t)}{\sum_l s_l(t)}(1+\delta_i)+
\sum_{j\in{\cal V}(i)}m\frac{s_j(t)}{\sum_l s_l(t)}
\delta_j \frac{w_{ij}(t)}{s_j(t)}\\
\frac{dk_i}{dt} &=&m\frac{s_i(t)}{\sum_l s_l(t)} \ ,
\label{eq_evol_sk_rand}
\end{eqnarray}
where now the explicit dependence from the specific $\delta_i$ of each
vertex is properly considered.
For each newly added edge from  the vertex $n$ to the existing vertex 
$i$, $\sum_j s_j$ is increased by $2+2\delta_i$. 
On the average it is therefore natural to consider that the total
strength is increasing linearly with time as 
\begin{equation}
\sum_{i=1}^t  s_i(t) \simeq 2m(1+\delta')t \ .
\label{eq:sum_s}
\end{equation}
We assume that $\delta'$ is a well defined constant, which is
certainly the case if $\rho(\delta)$ is bounded. It is worth remarking
that $\delta'\neq <\delta>$ since during the growth process, vertices
with larger $\delta_i$ will be preferentially chosen. The quantity
$\delta'$ will be determined self-consistently by the general
solution.

We use Eq.~(\ref{eq:sum_s}) in the evolution
Eqs.~(\ref{eq_evol_ss_rand}) and since $\sum_{j\in{\cal V}(i)}
w_{ij}\delta_j$ is a sum over the neighbors of $i$ {\em that
are chosen by the strength driven dynamics}, we assume that
$\delta_j\simeq\delta'$ and therefore $\sum_{j\in{\cal V}(i)} w_{ij}
\delta_j \simeq \delta' s_i$. We then obtain
\begin{eqnarray}
\nonumber
\frac{ds_i}{dt}&=&
\frac{\delta_i + \delta' +1}{2\delta' +2} \frac{s_i(t)}{t}\\
\frac{dk_i}{dt}
&=&\frac{s_i(t)}{2 (1+\delta') t} \ .
\label{eq_evol2_rand}
\end{eqnarray}
The integration of these equations with the initial conditions
$k_i(t=i)=s_i(t=i)=m$ yields
\begin{eqnarray}
\label{s_evol_rand}
s_i(t)&=&m ~\left(\frac{t}{i}\right)^{a_i} \\
k_i(t)&=&\frac{s_i(t) + m(\delta_i+\delta')}{\delta_i + \delta' +1}.
\label{k_evol_rand}
\end{eqnarray}
The strength and degree of the vertices therefore grow as
power-laws, with an exponent that depends on their ability to
redistribute weights:
\begin{equation}
a_i = \frac{1+\delta_i + \delta'}{2(1+\delta')} \ .
\label{ai}
\end{equation}
Equation (\ref{k_evol_rand} thus shows that also in the heterogeneous 
model the strength and
connectivity are proportional, with a coefficient depending on
$\delta_i$
\begin{equation}
s_i = k_i (1+\delta_i + \delta') - m (\delta_i + \delta') \ .
\end{equation}
The knowledge of the time behavior of $s_i(t)$ makes it possible to obtain
the probability distribution of connectivities and strengths. For example
the distribution of strengths can be written as 
\begin{equation}
  P_s(s, t) = \int d\delta \rho(\delta)
\frac{1}{t+N_0} \int_0^{t} \delta(s - s_i(t)) dt_i \ ,
\end{equation} 
where $\delta(s - s_i(t))$ is the Dirac delta function (not to be
confused with the heterogeneity parameter). Since $s_i(t) \propto
(t/t_i)^{a(\delta)}$ we obtain
\begin{equation}
P(s) \propto  \int d\delta \frac{\rho(\delta) }{a(\delta)}
 \left( \frac{1}{s}\right)^{1/a(\delta)+1} \ .
\end{equation}
which shows that the precise form of $P(s)$ depends on $\rho(\delta)$.
Finally, the proportionality of $k_i$ and $s_i$ ensures that their
distributions have the same form.

Along the same lines we can obtain the dynamical evolution 
of the weights. Indeed,  $w_{ij}$ grows each time
a new node connects to either $i$ or $j$, its evolution equation being
\begin{eqnarray}\nonumber
\frac{dw_{ij}}{dt}&=&m\frac{s_i}{\sum_ls_l}\delta_i\frac{w_{ij}}{s_i}
+m\frac{s_j}{\sum_ls_l}\delta_j\frac{w_{ij}}{s_j}\\
&=&  \frac{\delta_i + \delta_j}{2(1+\delta')}
\frac{w_{ij}}{t} \ .
\end{eqnarray}
This readily implies  that $w_{ij}(t) \propto t^{b_{ij}}$ with 
\begin{equation}
b_{ij}=\frac{\delta_i + \delta_j}{2(1+\delta')}.
\end{equation}

The behavior of the model depends explicitly on the value $\delta'$,
that has to be self-consistently determined. The consistency of the
solution is obtained using
Eqs.~(\ref{s_evol_rand},\ref{k_evol_rand},\ref{ai}) which give
\begin{eqnarray}\nonumber
\sum_{i=1}^t s_i(t) &\approx& \int d\delta \rho(\delta) 
\int_1^t dt_0 m \left( \frac{t}{t_0} \right)^{a(\delta)} \\
&=& m \int d\delta \rho(\delta) 
\frac{t -t^{a(\delta)}}{1-a(\delta)}
\label{eq:sums2}
\end{eqnarray}
with $a(\delta)=(1+\delta + \delta')/(2(1+\delta'))$. Since $s_i$
cannot grow faster than $t$, $a(\delta)$ has to be less than $1$ and
using Eq.~(\ref{eq:sum_s}), we obtain from Eq.~(\ref{eq:sums2}) in the
large time limit
\begin{equation}
2 m (1+\delta') = m \int \frac{d\delta \rho(\delta) }
{ 1 - \frac{1+\delta + \delta'}{2(1+\delta')} }
\end{equation}
or
\begin{equation}
\int \frac{d\delta \rho(\delta) }{1+\delta'-\delta} = 1 \ .
\end{equation}
which determines the value of $\delta'$.  Finally, we note that these
results are valid only if the quantity $\delta_i$ is bounded: if it
is not the case, the basic assumption that $\sum_i s_i$ grows linearly
is no longer true~\cite{Bianconi:2001}.

It is clear from the above solution that the graph's properties will
depend upon the particular form of the coupling distribution
$\rho(\delta)$. In order to compare with numerical simulations of the
model we analyze the specific case of a uniform distribution
$\rho(\delta)$ in the interval between $\delta_{min}$ and
$\delta_{max}$. The equation for $\delta'$ can be explicitly solved,
obtaining
\begin{equation}
\delta' = 
\frac{(\delta_{max}-1) \exp(\delta_{max}-\delta_{min}) +1 - \delta_{min}}
{ \exp(\delta_{max}-\delta_{min}) - 1}.
\label{deltap}
\end{equation}
We are therefore in the position to provide an explicit value of the
exponent $a(\delta)$ for the evolution of $s$ and $k$ during the
growth of the network.
\begin{figure}[t]
\vskip .5cm
\begin{center}
\epsfig{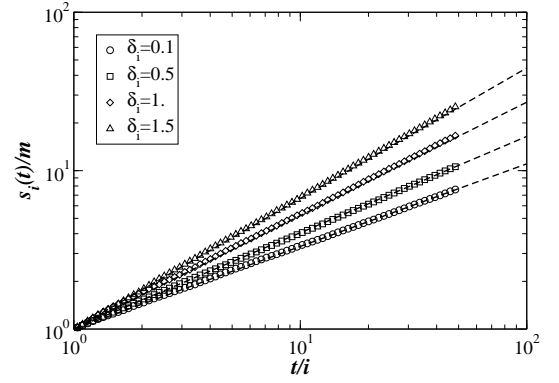}
\end{center}
\caption{Model with random $\delta_i$: evolution of $s_i/m$ for a given
  node $i$ with various values of its redistribution parameter
  $\delta_i$.  $m=2$, $N=10^4$, average over $1000$ networks with 
  the same realization of $\delta_j \in [0;2]$ ($j \ne i$). The dashed
  lines correspond to the predicted power-laws $(t/i)^{a(\delta_i)}$.} 
\label{fig:random1}
\end{figure}
\begin{figure}[t]
\vskip .5cm
\begin{center}
\epsfig{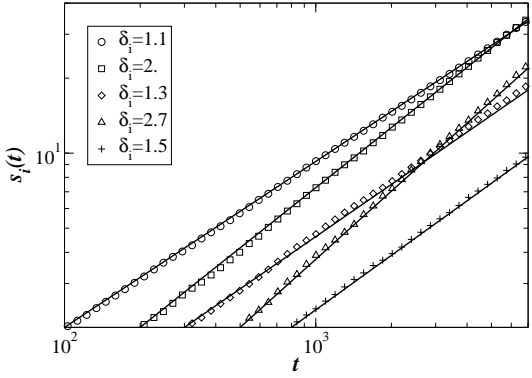}
\end{center}
\caption{Model with random $\delta_i$: evolution of $s_i$ for various
  nodes.  $m=2$, $N=10^4$. Symbols: results of numerical simulations, average
  over $1000$ networks with the same realization of $\delta_j \in [1;3]$.
  Nodes arriving later can overcome younger nodes. Lines are the prediction
  (\ref{s_evol_rand}) with the exponent given by (\ref{ai}) and
  (\ref{deltap}). }
\label{fig:random1bis}
\end{figure}
Similarly, the strength probability distribution can be written as
\begin{equation}
P(s) \propto \int_{a_{min}}^{a_{max}} \frac{da}{a} 
\left( \frac{m}{s} \right)^{1+1/a} \ ,
\end{equation}
whose behavior at  large $s$ is:
\begin{equation}
P(s)  \propto  \frac{1}{s^{1+1/a_{max}} \log (s)}
\label{ps_rand}
\end{equation}
where $a_{max}=(1+\delta_{max}+\delta')/(2(1+\delta'))$ is the largest
possible value of the exponent $a(\delta)$. 

A test of the analytical results is obtained by the direct inspection
of networks obtained by numerical simulations of the model in the case
of heterogenous coupling. In particular we consider networks
generated by uniform distribution of the coupling constants in
specific intervals as reported in the figure captions.  The striking
agreement of the analytical predictions
[Eqs.~(\ref{s_evol_rand},\ref{ai},\ref{deltap})] with the numerical
results is shown in Figs.~\ref{fig:random1} and \ref{fig:random1bis}.
It is worth noticing the remarkable agreement shown in these figures,
obtained without any free parameter.
\begin{figure}[t]
\vskip .5cm
\begin{center}
\epsfig{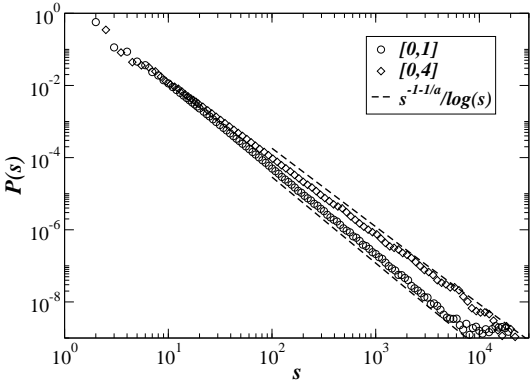}
\end{center}
\caption{Model with random $\delta_i$: distribution of strengths $P(s)$
  for uniform distributions $\rho(\delta)$, $\delta_j \in [0;1]$ and $\delta_j
  \in [0;4]$. Dashed lines are the theoretical predictions (\ref{ps_rand})
  $s^{-1-1/a_{max}}/\log(s)$. $N=10^5$, $m=2$, average over $1000$
  realizations.  }
\label{fig:randomps}
\end{figure}
\begin{figure}[t]
\vskip .5cm
\begin{center}
\epsfig{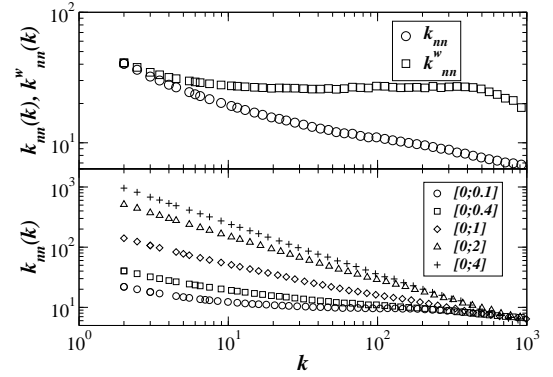}
\end{center}
\caption{$m=2$; Top: $k_{nn}(k)$ and $k_{nn}^w(k)$ 
for a uniform distribution $\rho(\delta)$ with 
$\delta_j \in [0;0.4]$; Bottom:
$k_{nn}(k)$ for uniform distributions $\rho(\delta)$
in various intervals $[\delta_{min},\delta_{max}]$.
}
\label{fig_knn_knnw_rand_delta}
\end{figure}
The statistical distributions of the quantities of interest are also in
good agreement with the analytical prediction as shown 
in Fig.~\ref{fig:randomps} for the strength distribution. Finally, 
as shown in Fig.~\ref{fig_knn_knnw_rand_delta}, the correlation and
clustering properties are non-trivial as well. As in the case of the
homogeneous coupling the average nearest neighbors degree and the
clustering coefficient have a clear structure with a
hierarchical ordering of high and low degree vertices. Also in this
situation the inclusion of weights in the characterization of
correlations provide additional information on the structure of the
network. While more cumbersome because of the heterogeneous nature of 
the coupling, a general understanding of the observed properties can
be obtained along the reasoning reported for the homogeneous coupling
model; in all cases, the dynamical growth process itself is at the
origin of the complex architecture and structure of the generated networks. 

\section{Non-linear coupling}
\label{sec:VI}

In the previous sections we  considered the coupling term $\delta_i$
as independent of the topological and weight properties of the vertex
$i$. We can however think of different situations in which the 
perturbation depends on the centrality of the node as measured by 
its strength or degree. In the airline network, for example, 
this might mimic the fact that larger is the  airport and larger 
is the increase of traffic with a much larger response to the creation 
of a new connection compared to a small airport. 
It is thus natural to investigate the consequence of 
non-linear coupling forms.

The  simplest non-linear coupling consists in 
considering that $\delta_i$  is proportional to $s_i$. 
In order to avoid unrealistic divergences a cut-off $s_0$
is however needed to bound the coupling, leading to the reinforcement
rule
\begin{equation}
\Delta w_{ij}=\delta\frac{w_{ij}}{s_i} 
\left[ s_0 \tanh \left( \frac{s_i}{s_0} \right) \right]^a \ .
\end{equation}
This relation simply expresses that the larger the traffic on the
vertex $i$ and the larger will be the traffic attracted to it during
the weights' dynamics. The total change in $s_i$ when a link is added
is now $\delta [ s_0 \tanh (s_i/s_0) ]^a$. For strength smaller than
the cutoff $s_0$, this change grows as $s_i^a$, and saturates to
$\delta s_0^a$ for very large strengths.

\begin{figure}[t]
\vskip .5cm
\begin{center}
\epsfig{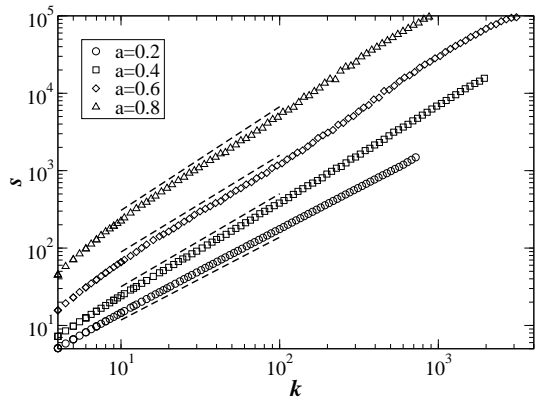}
\end{center}
\caption{$N=5000$; $s_0=10^4$. $s$ vs $k$ for $a=0.2, 0.4, 0.6,
0.8$. Dashed lines have slope $1.07$, $1.21$, $1.25$ and $1.34$ (from
bottom to top). For values of $s$ larger than $s_0$ there is a
crossover towards $\beta=1$.  }
\label{fig:th_Sa}
\end{figure}

The nonlinear mechanism makes the analytical solution very difficult
to find and we rely on a numerical study of the model to inspect its
topological and weight properties. We find that $\sum s_i(t)$ now
grows faster than $t$, seemingly like $\exp(t^b)$. Analogously, we
observe that vertices' degree and the strength grow faster than simple
power-laws.  Very interestingly, we observe that the strength grows as
a power-law with the connectivity, $s \sim k^\beta$ with $\beta > 1$,
as shown from the numerical simulations reported in
Fig.~\ref{fig:th_Sa}. The exponent $\beta$ increases with $a$ but it
is independent from $\delta$. This result raises the possibility that
some real-world networks, in which a value $\beta>1$ is observed, are
governed by local non-linear reinforcement processes. This could be
the case for the airport network where $\beta\simeq 1.5$ is
observed~\cite{Barrat:2004a}.

\begin{figure}[th]
\vskip .5cm
\begin{center}
\epsfig{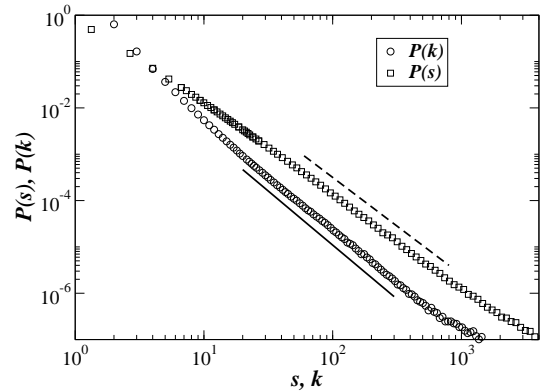}
\end{center}
\caption{$N=5000$; $s_0=10^4$. $P(s)$ and $P(k)$ for $a=0.8$, $\delta=0.1$. 
The data for $P(k)$ have been shifted vertically for clarity.
Continuous and dashed lines correspond to the power laws $k^{-2.33}$
and $s^{-2.1}$ respectively.}
\label{fig:pspknonlin}
\end{figure}

An interesting consequence is then observed for the degree and
strength probability distribution. While both distributions still
behave as power laws, $P(k)\sim k^{-\gamma_k}$ and $P(s)\sim
s^{-\gamma_s}$, as shown in Fig.~\ref{fig:pspknonlin}, they exhibit
different exponents $\gamma_s$ and $\gamma_k > \gamma_s$ in contrast
with all the situations considered previously where we found
$\gamma_s=\gamma_k$. This is obviously linked to the fact that
$\beta\neq 1$ and it is not difficult to show that
\begin{equation}
\gamma_s=\frac{\gamma_k}{\beta}+\frac{\beta-1}{\beta}.
\end{equation}
The weight distribution $P(w)$ is also power-law distributed and in
addition, all distributions get broader as either $a$ or $\delta$ are
increased. Finally, we note that the correlations and clustering
properties exhibit also in this case non-trivial spectrum as a
function of the degree $k$, signalling the presence of a hierarchical
architecture also in the presence of a non-linear coupling.

It is clear that the results obtained for non-linear coupling
mechanisms are depending upon the detailed form of the coupling and
further studies are needed in order to understand the variations of
time behavior and distribution exponents as a function of the various
parameters defining the reinforcement dynamics. Obviously, a detailed
study of all non-linear coupling mechanisms is impossible and each
modeling effort must be driven by specific insights on the dynamics of
the real systems under examination.

\section{Conclusions}
\label{sec:VII}

In this paper we have presented a general model of growing networks 
that considers the effect of the coupling between topology and
weights dynamics. We investigated in details several 
coupling mechanisms including the effect of randomness and 
non-linearity in the redistribution process.

The model produces graphs which display non trivial complex 
and scale-free behavior that depend on the detailed coupling form. 
In particular, different quantities such as strength,
degree and weights are distributed according to 
power-laws with exponents which are not universal and depend on 
the specific parameters that control the local microscopic weights' 
dynamics. This result hints to a simple explanation of the lack 
of any universality observed in real-world networks.
In addition, the dynamics generates spontaneously a
non-trivial architecture in which nodes with different degrees are
arranged in a hierarchical way as indicated by the clustering and
correlation properties measured in the obtained networks.

While  many  other parameters and dynamical features 
may be entering the dynamics of 
real-world networks, we believe that the present model might provide 
a general starting point for the realistic modeling of several systems
where the interplay of topology and  traffic is a key point in the
determination of the global network's properties.

\begin{acknowledgments}
  We thank R. Pastor-Satorras and M. Vergassola for useful comments on the
  manuscript.  A.B and A. V. are partially funded by the European Commission -
  Fet Open project COSIN IST-2001-33555 and contract 001907 (DELIS).
\end{acknowledgments}




\end{document}